\documentclass[twocolumn]{aastex7}

\usepackage{epstopdf}
\newcommand\mps{Max-Planck-Institut f\"ur
Sonnensystemforschung, Justus-von-Liebig-Weg 3, 37077 G\"ottingen, Germany}
\newcommand\kfu{Institute of Physics, University of Graz, 8010 Graz, Austria}
\begin{document}

\title{Molecular hydrogen controls the temperatures of flares on TRAPPIST-1}

\author[0000-0002-8842-5403]{Shapiro, A.I.}
\affiliation{\kfu}
\affiliation{\mps}
\email[show]{alexander.shapiro@uni-graz.at}  

\author[0000-0002-6087-3271]{Kostogryz, N.}
\affiliation{\mps}
\email[]{kostogryz@mps.mpg.de}

\author[0000-0002-6892-6948]{Seager, S.}
\affiliation{Department of Physics and Kavli Institute for Astrophysics and Space Research, Massachusetts Institute of Technology, Cambridge, USA}
\affiliation{Department of Earth, Atmospheric and Planetary Sciences, Massachusetts Institute of Technology, Cambridge, USA}
\affiliation{Department of Aeronautics and Astronautics, Massachusetts Institute of Technology, Cambridge, USA}
\email[]{} 

\author[0000-0002-0929-1612]{Witzke, V.}
\affiliation{\kfu}
\email[]{} 

\author[0000-0003-2415-2191]{de Wit, J.}
\affiliation{Department of Earth, Atmospheric and Planetary Sciences,\\ Massachusetts Institute of Technology, Cambridge, MA 02139, USA}
\email[]{}

\author[0009-0009-3020-3435]{Vasilyev, V.}
\affiliation{\mps}
\email[]{}  

\author[0000-0003-2073-002X]{Astrid M. Veronig}
\affiliation{\kfu}
\affiliation{University of Graz, Kanzelh\"ohe Observatory for solar and Environmental Research, Kanzelh\"ohe 19, 9521 Treffen, Austria
}
\email[]{}

\author[0000-0002-6568-6942]{Robert~Cameron}
\affil{\mps}
\email[]{}  

\author[0000-0001-9921-0937]{Hardi~Peter}
\affil{\mps}
\affiliation{Institut für Sonnenphysik (KIS), Georges-Köhler-Allee 401a, 79110 Freiburg, Germany}  
\email[]{}  

\author[0000-0002-3418-8449]{Solanki, S.K.}
\affiliation{\mps}
\email[]{}

\begin{abstract}
Early JWST observations of TRAPPIST-1 have revealed an unexpected puzzle: energetic white-light flares ($\rm{E} > 10^{30}$ erg) reach temperatures of only ${\sim}$3500--4000\,K, nearly three times cooler than typical solar flares, which peak around 9000--10000\,K. Here we explain this difference by identifying the physical mechanism that regulates flare temperatures on late M-dwarfs. The key factor is that in the cool, dense atmosphere of TRAPPIST-1, magnetic heating is strongly moderated by the dissociation of molecular hydrogen (H$_2$) into atomic hydrogen. This "H$_2$ dissociation thermostat" acts as an efficient energy sink, preventing flare regions from heating above ${\sim}4000$\,K. Our chemical equilibrium and heat capacity calculations show that this effect depends sensitively on stellar atmospheric pressure and the local abundance of H$_2$. In hotter stars, from early M dwarfs to solar-type stars, the scarcity of molecular hydrogen renders this mechanism ineffective; instead, atomic hydrogen ionization limits flare temperatures near ${\sim}$9000\,K.
\end{abstract}


\section{Introduction} \label{sec:intro}
Stellar magnetic activity has long been recognized as a fundamental process shaping the  environments of stars like our Sun. Systematic monitoring of magnetic activity in stars beyond the Sun began over half a century ago with Olin Wilson’s HK Project, which monitored magnetically driven variations in the Ca II H and K emission lines in solar-type stars \citep{Wilson1978}. In the decades that followed, sustained observational and theoretical efforts led to major progress in understanding stellar magnetic activity and its diverse manifestations \citep[see, e.g.,][for a review]{Basri2021}.

Despite its maturity, the field of stellar activity has witnessed a remarkable surge in discoveries over the past decade. Interestingly, it happened largely thanks to the missions and programs whose primary goal was not to study stars themselves, but to characterize the exoplanets they host. Direct observations of exoplanets remain limited, meaning that much of exoplanetary science currently depends on precise measurements of host stars. Alongside planetary signals, these observations often reveal signatures of stellar magnetic activity. While such stellar “contamination” presents a major challenge for exoplanet research \citep[see, e.g.,][and references therein]{RV_final_report, SAG2023}, it also provides an unprecedented opportunities to study stellar activity—often leading to unexpected results and driving revisions to existing models.

\begin{figure*}
    \centering
    \includegraphics[width=0.95\linewidth]{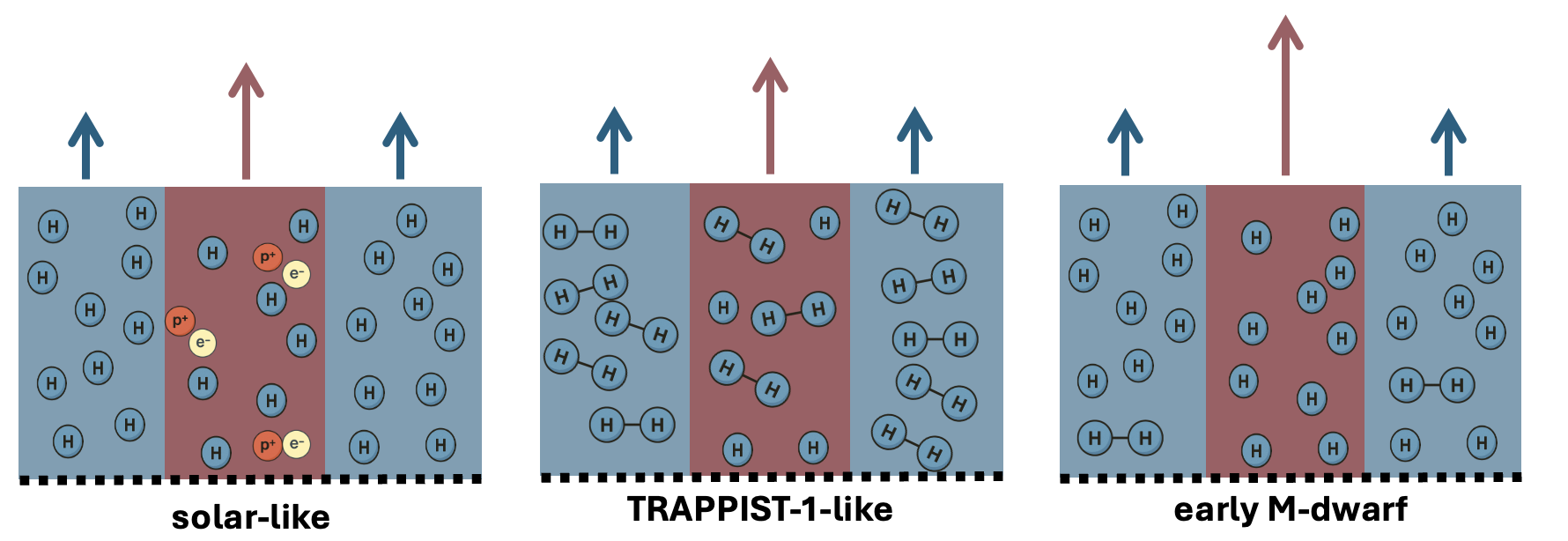}
    \caption{ {\bf Illustration of flare heating in three types of stars.}  The dashed line marks the stellar surface; blue areas show the quiet atmosphere, and red areas show regions heated during a white-light flare. Vertical arrows indicate the radiative output, with length proportional to intensity. In solar-like stars (left), heating is regulated by hydrogen ionization, which limits temperatures of white-light flares to $\sim 9000$ K. In late M dwarfs like TRAPPIST-1 (middle), molecular hydrogen dissociation acts as a thermostat, capping temperatures of white-light flares near $\sim 3500-4000$ K. In early M dwarfs (right), insufficient concentration of $\rm H_2$ means the dissociation thermostat is ineffective, allowing higher white-light flare temperatures. }  
    \label{fig:cartoon}
\end{figure*}

Here, we address one of such cases brought by the James Webb Space Telescope (JWST) observations of cool M8-dwarf TRAPPIST-1. It is currently one of the most intensively studied main-sequence stars after the Sun, owing to the seven rocky planets it hosts \citep[][]{Gillon2017, Luger2017}. Despite substantial efforts, no definitive evidence for atmospheres around its planets has been found to date. One of the possible reasons for this is a notorious magnetic activity of TRAPPIST-1. It manifests itself through variability in the transmission spectra of TRAPPIST-1’s planets \citep[][]{Lim2023, Radica2025} and frequent energetic flares, which have become a ubiquitous aspect of TRAPPIST-1 transit observations \citep[][]{Howard2023}. The energetic flares ($\rm{E}>10^{30}$ erg) occur on TRAPPIST-1 at a rate of $\sim$1000/year. This is as frequent as on the Sun, despite the Sun being nearly 2000 times more luminous than TRAPPIST-1.

The very detection of stellar flares with JWST is not, in itself, surprising — stellar flares on main-sequence stars have been known for almost a century \citep{first_flare}. More recently, flares have been observed on thousands of stars by the Kepler and TESS missions \citep[][]{Maehara2012, Notsu2019, Gunther2020, Valera2024}. These large surveys have revealed numerous properties of stellar flares, including their occurrence rates across different spectral types and stellar ages.  However, the white-light nature of Kepler and TESS observations did not allow for direct measurements of one of the most fundamental properties of white-light flares — their temperature. Knowing the flare temperature is essential for estimating the bolometric energy output of these events. In the absence of direct spectral measurements, flare temperatures have typically been assumed to match those of solar white-light flares, around 9000–10,000 K \citep[][]{Kretzschmar2011}.

The JWST’s spectral resolution and sensitivity brought an unprecedented ability to study flares spectra and hence measure temperatures. 
\cite{Howard2023} used  observations by Near Infrared Spectrograph (NIRSpec/JWST) and Near Infrared Imager and Slitless Spectrograph (NIRISS/JWST) to {show that the effective temperatures of flares on TRAPPIST-1 outside of the short impulsive peak are almost three times lower than those of solar white-light flares (3500–4000 K vs. solar 9000–10000 K in the visible). In the following we refer to this value during the slowly decaying phase as the flare temperature. Its low value contradicts  the commonly adopted “solar paradigm”  of 9000 K white-light flares}.

 {In this Letter we seek to explain what sets this unexpectedly low flare temperature on TRAPPIST-1. Rather than attempting a full radiative-hydrodynamic (RHD) treatment of the flare, we focus on the slowly decaying, “cool component” that dominates the optical/infrared emission and JWST transit contamination during TRAPPIST-1 flares \citep[][]{Kowalski2013, Howard2023}. Our approach is deliberately simple: we use the chemical-equilibrium module of the MPS-ATLAS code \citep[][]{MPS-ATLAS} to compute hydrogen concentrations and isobaric heat capacities atmospheric layers representative of TRAPPIST-1. We combine these with back-of-the-envelope energy-budget arguments. While do not solve the time-dependent hydrodynamics or non-Local Thermodynamic Equilibrium radiative transfer, we identify a robust thermodynamic mechanism that detailed RHD or magnetohydrodynamic (MHD) model must satisfy. We show that in the high-pressure photosphere of TRAPPIST-1, magnetic heating deposited by a flare is efficiently moderated by the dissociation of molecular hydrogen, which acts as a powerful thermostat. This “$\rm H_2$  dissociation thermostat” keeps  TRAPPIST-1 flare temperatures well below those of hotter stars, where the scarcity of $\rm H_2$  renders this mechanism ineffective.}

 %{\bf Our approach is based on a  RHD models such as RADYN have been essential for understanding how energy released by magnetic reconnection is transported and deposited in stellar atmospheres. These models explicitly capture key processes, including electron‐beam energy deposition, conductive and radiative heating. They also solve the time-dependent equations of hydrodynamics together with non-equilibrium radiative transfer for a vertical column of the atmosphere, basically doing everything that the referee suggested. 
%}

\section{Hydrogen Thermostat Effect in Stellar Atmospheres}
In Sect.~\ref{subsect:H} we revisit the concept of the hydrogen ionization thermostat that limits the temperature of solar white-light flares to $\sim$9000~K. Modern solar flare models are highly advanced and account for numerous non-trivial effects that cause deviations from a simple 9000~K Planck spectrum. As a result the concept of the hydrogen ionization thermostat has largely disappeared from literature.  Nevertheless, we briefly recall it here, as it provides the foundation for understanding our main result: a similar thermostat mechanism controls the relatively low flare temperatures observed on TRAPPIST-1. In Sect.~\ref{subsect:H2}, we introduce this mechanism: the $\rm H_2$ dissociation thermostat. It operates in cool atmospheres of late M-dwarfs and plays an analogous role to the hydrogen ionization thermostat in solar flares.  We demonstrate that this $\rm H_2$ thermostat constrains the flare temperatures on TRAPPIST-1.

\subsection{Hydrogen Ionization as a Thermostat in the Atmospheres of Solar-like Stars}\label{subsect:H}
Solar flares are powered by magnetic reconnection in the corona. Rapid reconfiguration of the coronal magnetic field during reconnection converts stored magnetic energy into kinetic and thermal energy of plasma \citep[see the seminal reviews by][]{Priest2002, Shibata2011LRSP}. A portion of this energy propagates outward
%upward 
into the heliosphere being transferred by high-speed plasma outflows (jets), shocks, and possibly coronal mass ejections (CMEs). These upward-directed phenomena have been intensively studied due to their central role in space weather and the growing interest in the possibility of extreme solar events capable of impacting our daily life \citep[see recent review by][]{Usoskin2023SSR}.

Another part of the energy released during the flare %magnetic 
reconnection 
propagates downward, being transferred primarily by non-thermal particles and magnetohydrodynamic waves \citep[see Chapter 5 of][for a comprehensive review of proposed transport mechanisms]{Shibata2011LRSP}. This energy passes through the solar corona and chromosphere and %reaches 
part of it may also reach the photosphere. 
In the standard flare model, a significant fraction of the released energy goes into the acceleration of non-thermal electrons. They transverse the tenuous corona with rather unaltered spectra, and deposit their energy predominantly in the chromosphere (which acts as a thick-target), by Coulomb collisions with ambient electrons \citep{Brown1971}. The resulting impulsive heating and associated pressure gradient causes hot plasma to expand upward into the coronal loops in a process called chromospheric evaporation \citep{Fisher1985}. These flaring loops then emit enhanced extreme ultraviolet (EUV) and X-ray emission, which
%The energy deposited in the corona and upper chromosphere is subsequently released as radiation, leading to enhancements in extreme ultraviolet (EUV) and X-ray emission. 
%These enhancements 
are readily detectable against the relatively weak EUV and X-ray background emission of the quiet Sun and, therefore, serve as primary tracers of flares \citep{Fletcher2011SSRv, Benz2017LRSP}. The %propagation 
transport of flare energy into the temperature minimum region and photosphere leads to a brightening in the visible continuum, known as white-light flares \citep[][]{Neidig1989}, which are generally associated with the strongest solar flares. The detailed processes of how the white-light emission in (solar) flares is produced, is still under debate. Electron beams are not able to penetrate deep enough to directly heat the photosphere. Alternatively, the heating there may result from radiative back-warming  of Balmer and Paschen continua \citep{Allred2006} or chromospheric condensations \citep{Kowalski2015}.

\begin{figure*}
    \centering
    \includegraphics[width=0.99\linewidth]{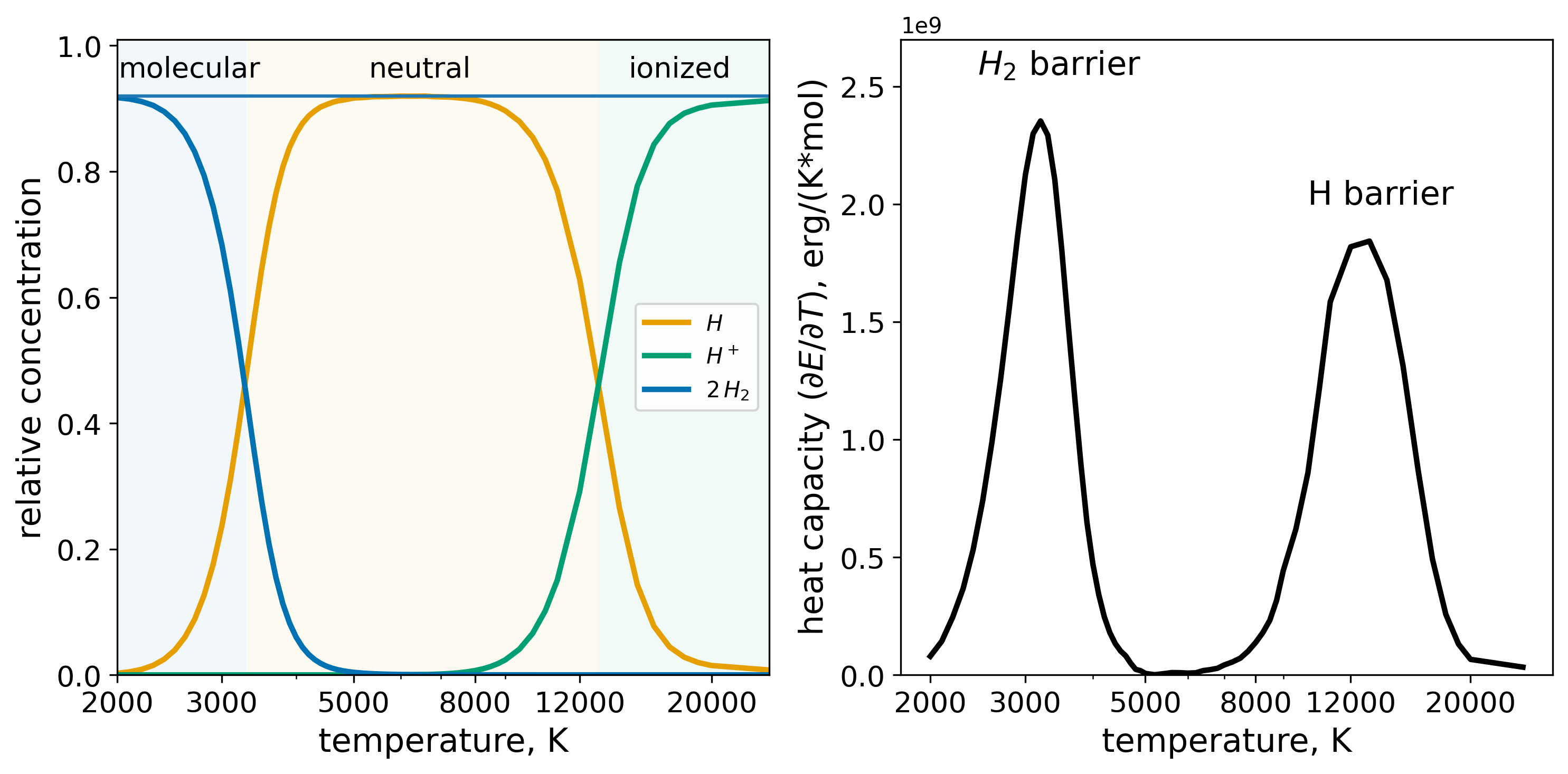}
    \caption{ {\bf Molecular hydrogen dissociation and atomic hydrogen ionization thermostats in stellar atmospheres}.  Left panel: Relative concentrations of molecular, atomic, and ionized hydrogen, normalized to the total concentration of heavy particles (i.e., all particles except electrons), shown as functions of temperature. The concentration of molecular hydrogen is multiplied by two to account for the fact that each molecule consists of two hydrogen atoms.  The horizontal blue line indicates the condition where hydrogen exists entirely in a single stage, either molecular, neutral, or ionized. Right panel: isobaric heat capacity. The calculations are performed for solar elemental composition and pressure, $\sim 0.1$ bar, typical for the solar surface. The key message of the figure is that both dissociation and ionization occur over a very narrow temperature range, resulting in two distinct peaks in heat capacity — the $\rm H_2$ and H thermostats.}
    \label{fig:diss:ion}
\end{figure*}

\begin{figure*}
    \centering
    \includegraphics[width=0.99\linewidth]{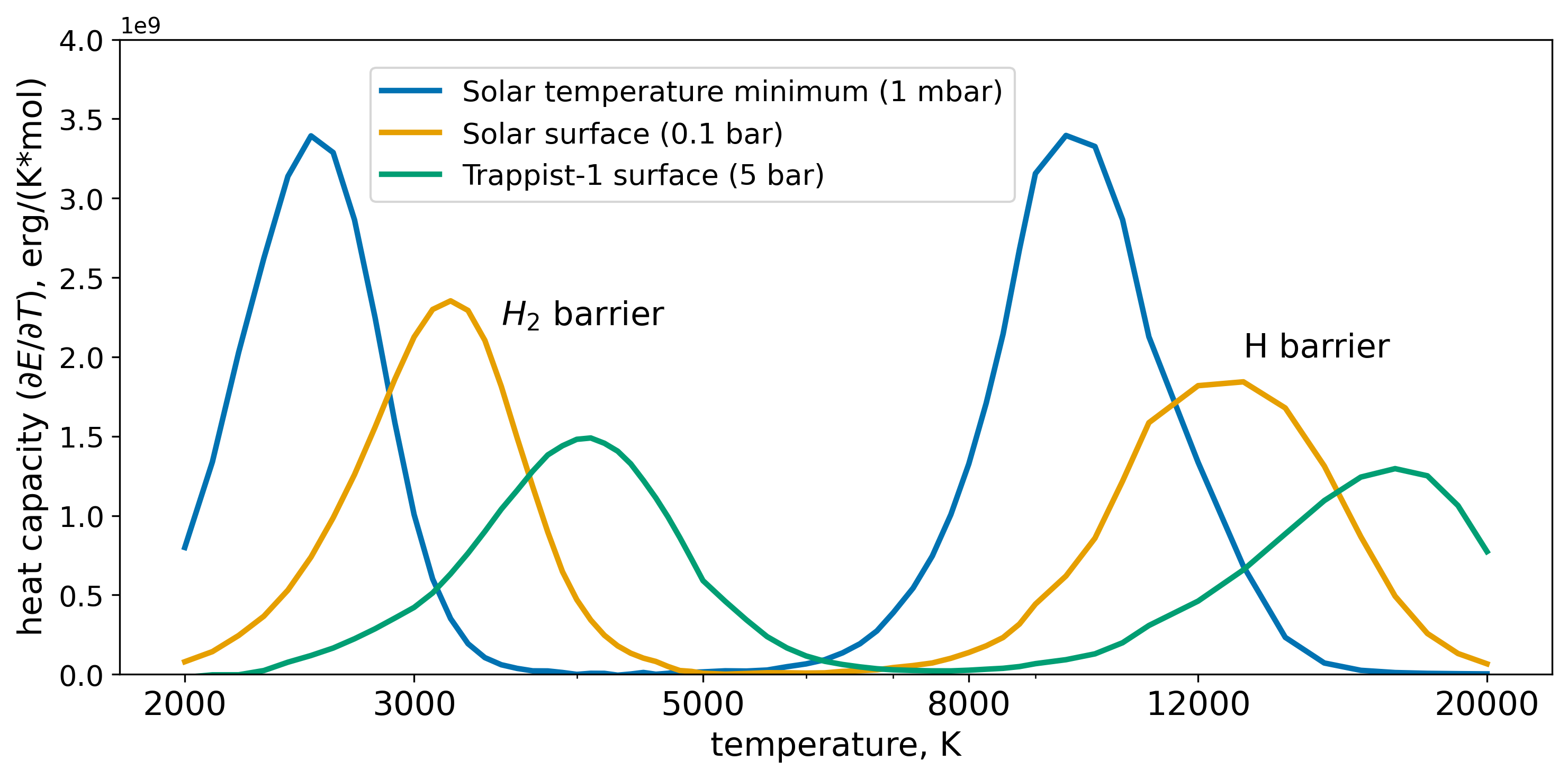}
    \caption{ {\bf Hydrogen thermostats at different pressure values.} Shown are heat capacities computed for solar elemental composition and pressure values of 1 mbar (typical for the solar temperature minimum region), 0.1 bar (typical for the solar surface), and 5 bar {(typical for TRAPPIST-1 optical surface, see Fig.~\ref{fig:TP_structure})}. The key message of the figure is that the thermostat temperatures depend on pressure, and therefore vary with the type of star and the height in the stellar atmosphere from which the white-light flare emission originates.}
    \label{fig:P_dep}
\end{figure*}

White-light flares have been known since the mid-19th century, when the first such event was visually observed by Richard Carrington  \citep{Carrington1859MNRAS}. Despite this early discovery, accurate measurements of the spectral energy distribution in white-light flares is challenging due to the strong  background flux from the quiet Sun \citep[see, e.g.,][]{Hudson2006SoPh, Heinzel2017ApJ} and became possible only with the advent of high-precision space-borne solar irradiance monitoring.  \citet{Kretzschmar2011} established observationally that the visible continuum emitted during solar white-light flares is well approximated by a blackbody with an effective temperature of around 9000 K. % based on a superposed epoch analysis of total and spectral solar irradiance  data from the SOHO and GOES spacecrafts. 

The existence of the temperature plateau around 9000--10000 K has been independently obtained in radiative-hydrodynamic models of flares, where it arises naturally due to the ionization of hydrogen. In these models, energy deposited by non-thermal electrons is efficiently absorbed by the hydrogen-rich stellar atmosphere, where much of it is consumed by ionizing hydrogen atoms (and consequent radiative cooling) rather than raising the local temperature \citep[e.g.,][]{Hawley1992, Allred2005, Kowalski2015}. Interestingly, a similar $\sim$7000 K plateau is present in many semi-empirical 1D models of the quiet solar chromosphere  \citep[see, e.g. the often referenced Figure 1 from][]{VAL1981ApJS}. The chromospheric plateau in these semi-empirical models happens at lower pressure values than hydrogen ionization due to solar flares and, thus, correspond to a lower temperature.

We note that the 9000 K concept for %the 
white-light solar flares is only an approximation to a more complex spectrum. For example, color temperatures inferred from UV fluxes are substantially higher than those derived from the visible, both for solar and stellar flares \citep[see, e.g.,][]{Fletcher2007, Howard2020, Kowalski2024LRSP}. Recently, \citet{Berardo2025} confirmed this relation between the UV and white-light temperatures for TRAPPIST-1 flares using HST/STIS observations. {Simultaneous ULTRACAM–spectroscopic observations also reveal that the optical/NUV continuum is intrinsically two-component, consisting of a hot Paschen‐dominated blackbody-like emission plus an excess Balmer recombination continuum that strengthens toward the decay phase \citep[][]{Kowalski2016}.}
The effort to accurately model this spectrum has reached a certain degree of maturity reproducing many observations \citep[][]{Kowalski2024LRSP}.

Along the same line the 7000 K chromospheric plateau is much less pronounced if the non-equilibrium effects are taken into account in the chromospheric models \citep{Golding2016ApJ}. Nevertheless, a flare temperature of 9000\,K has been used not only for the Sun, but has also become a standard assumption for other stars. {In practice, this value is routinely adopted when converting flare amplitudes measured in broad photometric passbands, such as those of Kepler and TESS, into estimates of the bolometric radiative energy} 
\citep[see, e.g.,][]{Shibayama2013, Notsu2019, Gunther2020, Valera2024}.

\subsection{Hydrogen Dissociation as a Thermostat in the Atmospheres of Late M Dwarfs}\label{subsect:H2}
The analysis of four TRAPPIST-1 white-light flares by \cite{Howard2023} revealed flare temperatures very different from the solar values of 9000 K routinely assumed in the literature. During the impulsive phase of TRAPPIST-1  flares (defined as the period of rapid radiative output increase) the temperature briefly rised to approximately 5000 K. Within a few minutes, it decreased to 3000–3500 K, where it remained stable for 15--20 minutes \citep[see Figure 7 from][]{Howard2023}. 
% AV: I think it is Fig. 7 not 8 that is meant

We propose that  the stabilization of the TRAPPIST-1 flare temperatures at 3000–3500 K  reflects the action of the hydrogen dissociation thermostat. In the solar case, heating of the atmospheric layers where white-light flares originate is limited beyond 9000 K, as the energy input is efficiently absorbed by hydrogen ionization (left panel of Figure~\ref{fig:cartoon}).  In the cooler TRAPPIST-1 atmosphere, a similar regulatory effect occurs at lower temperatures, with energy absorption dominated by molecular hydrogen dissociation (middle panel of Figure~\ref{fig:cartoon}). The initial temperature rise above 3500 K may be attributed either to complex processes directly associated with magnetic reconnection and downward energy transport in the TRAPPIST-1 atmosphere, or to complete dissociation of hydrogen in the region where the white-light flare originates.

The key element of our proposed explanation is that the transition from predominantly molecular to predominantly atomic hydrogen occurs within a relatively narrow temperature range of approximately 3000--3500 K. This  is illustrated in the left panel of Figure~\ref{fig:diss:ion}, which shows the relative abundances of molecular, atomic, and ionized hydrogen as a function of temperature. %calculated for pressures typical of the solar surface. 
The sharp nature of the transition from molecular hydrogen to atomic  implies that even a modest temperature increase within the 3000--3500 K range leads to the dissociation of a significant fraction of hydrogen molecules, requiring substantial energy input. A similar regulatory effect occurs during hydrogen ionization, consistent with the analogous mathematical form of the Saha equation for ionization and the Guldberg–Waage law for chemical equilibrium.

The substantial energy required for the rapid dissociation of molecular hydrogen produces a pronounced peak in the heat capacity at around 3000~K (right panel of Figure~\ref{fig:diss:ion}). A similar peak appears at approximately 12000~K due to hydrogen ionization. This temperature is higher than the 9000~K discussed in Sect.~\ref{subsect:H} because it is calculated for the pressure at the solar surface.  The temperatures of thermostats depend on pressure, with lower pressures leading to lower characteristic temperatures (Figure~\ref{fig:P_dep}). Solar white-light flares originate higher in the atmosphere, near the temperature minimum region, where the pressure is significantly lower. 

The calculations shown in Figures~\ref{fig:diss:ion}–\ref{fig:P_dep} were performed using the chemical equilibrium module of the MPS-ATLAS code \citep{MPS-ATLAS}, adopting the solar elemental composition from \citet{Asplund2009}. {MPS-ATLAS solves for chemical equilibrium taking into account all essential diatomic molecules and their radicals, as well as the most important (for chemical equilibrium) polyatomic molecules  (including $\mathrm{H_2O}$, $\mathrm{CO_2}$, $\mathrm{HCN}$, $\mathrm{HNO}$, etc). It utilizes pre-tabulated equilibrium constants from the NIST–JANAF tables \citep[][]{Chase1998}, and dissociation energies from \citet{HuberHerzberg1979}. A detailed description of the chemical–equilibrium module is provided in \citet{MPS-ATLAS}. We note, however, that the curves in Figures~\ref{fig:diss:ion}–\ref{fig:P_dep} are governed primarily by the chemical balance between atomic and molecular hydrogen and are only weakly sensitive to the particular set of included species or to the adopted elemental abundances. The main contribution of helium and heavier elements is to reduce the maximum relative hydrogen abundance in the left panel of Figure~\ref{fig:diss:ion} to values below unity (as indicated by the horizontal blue line). }

\begin{figure}
    \centering
    \includegraphics[width=0.99\linewidth]{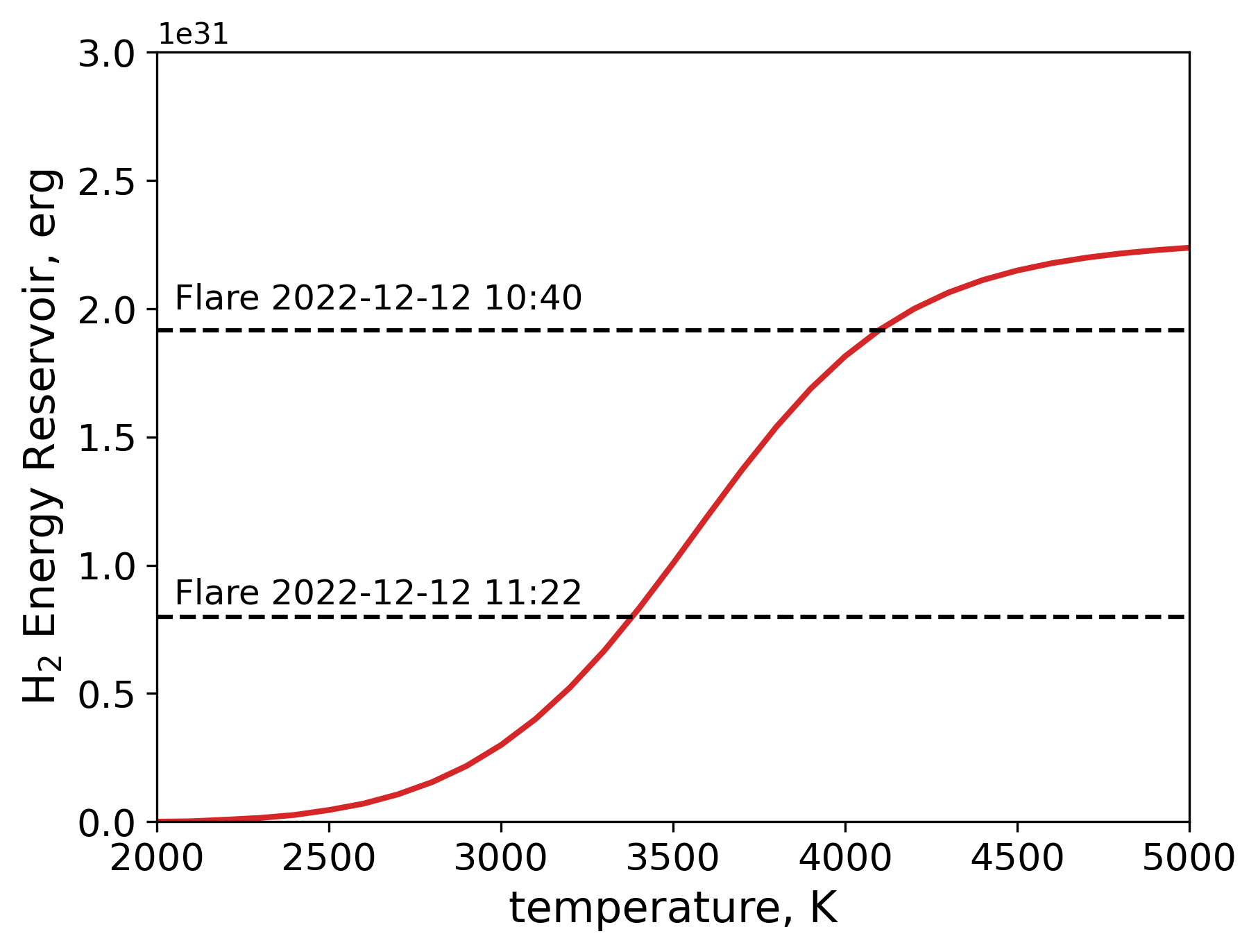}
    \caption{ {\bf The $\rm H_2$ energy reservoir.} Energy required to isobarically heat a mass of gas equivalent to 0.45\% of TRAPPIST-1’s atmosphere (typical of strong flares, see text for details) from 2000 K. The two dashed horizontal lines indicate bolometric energies of the two strongest TRAPPIST-1 flares analyzed by \cite{Howard2023}.  The figure demonstrates that the dissociation energy stored in $\rm H_2$ is sufficient to account for the flare’s total bolometric emission. }
    \label{fig:energy}
\end{figure}

The energy stored in the TRAPPIST-1 molecular hydrogen ('$\rm H_2$ energy reservoir') is sufficient to hamper the temperature rise of TRAPPIST-1 flares. 
We illustrated this by calculating the energy required to heat the portion of the atmosphere lying above an area that covers 0.45\% of the TRAPPIST-1 surface \citep[corresponding to the filling factors of the two strongest flares deduced by][]{Howard2023}.  Heating this region from 2000~K to 2500~K requires approximately $5 \times 10^{29}$ erg, while increasing the temperature from 3000~K to 3500~K demands more than an order of magnitude higher input, around $7 \times 10^{30}$ erg (Figure~\ref{fig:energy}). Above 4500~K, the required energy plateaus, reflecting the complete dissociation of molecular hydrogen.

The energies of the two strongest flares observed by \citet{Howard2023} remain below the threshold for full dissociation (see Figure~\ref{fig:energy}). It implies that the hydrogen dissociation thermostat has the potential to limit 
temperatures of most white-light flares on TRAPPIST-1. This argument can also be viewed from another perspective: association of hydrogen releases enough latent energy to account for the bolometric output of TRAPPIST-1 flares. This suggests that a substantial fraction of the energy deposited into the lower atmosphere of TRAPPIST-1 during the impulsive phase can be efficiently stored in the form of dissociated hydrogen, which later recombines and re-emits the energy after the direct flare heating subsides.

\begin{figure}
    \centering
    \includegraphics[width=0.99\linewidth]{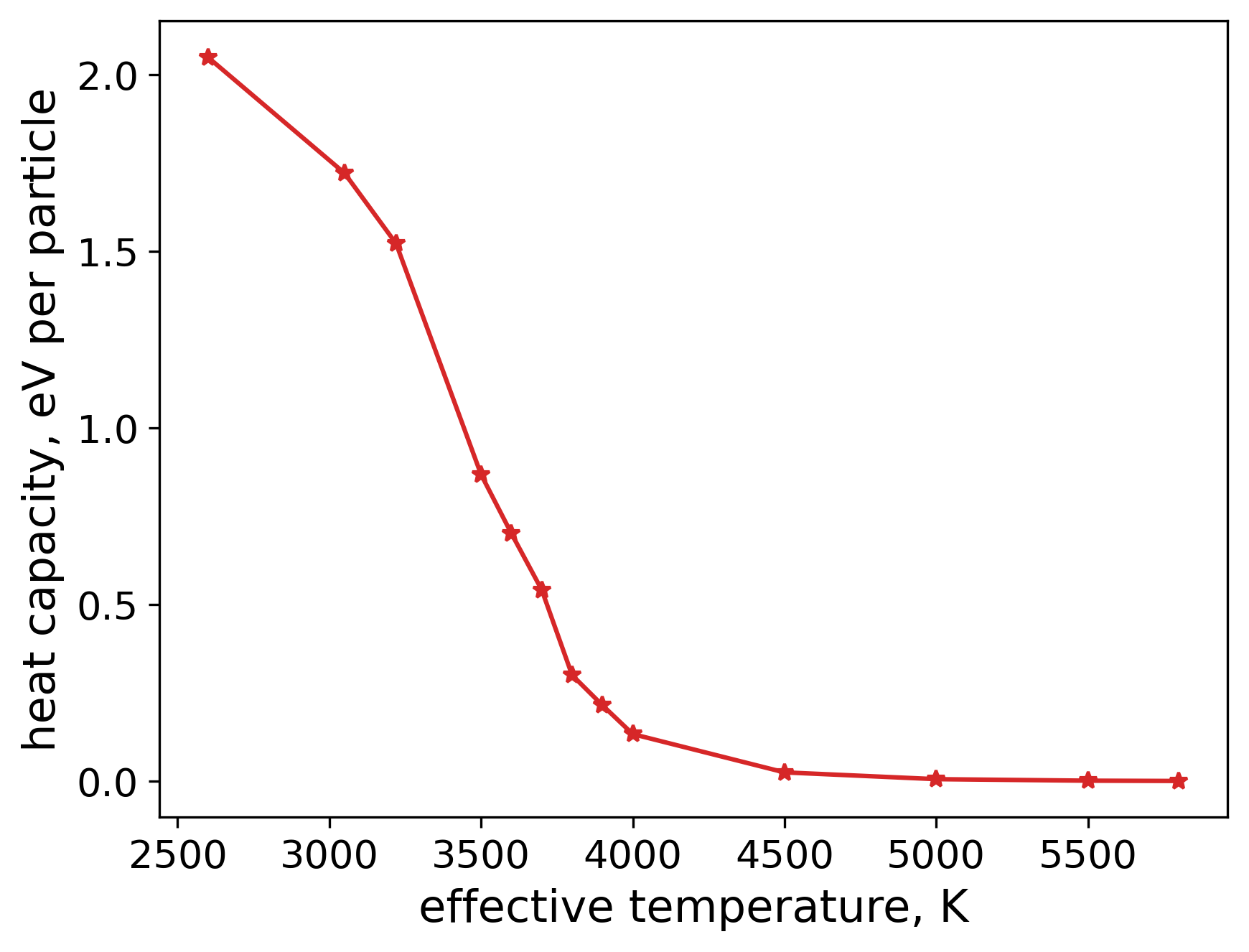}
    \caption{ {\bf The $\rm \bf H_2$ thermostat disappears for hotter stars}. The heat capacity (per particle) as a function of stellar effective temperature. The key takeaway from the figure is that the $\rm H_2$ thermostat weakens with increasing stellar effective temperature and essentially vanishes in early M- and late K-dwarfs. }
    \label{fig:hotter}
\end{figure}

The calculations shown in Figure~\ref{fig:energy} were performed for isobaric heating at a pressure equal to half of the TRAPPIST-1 surface value. We assumed that the entire atmospheric column is heated from 2000 K to a given target temperature. This may appear contradictory, considering that much of the TRAPPIST-1 atmosphere already exceeds 2000 K  \citep[recall that the effective temperature of TRAPPIST-1 is around 2600 K, see][]{Davoudi2024}. However, since heating from 2000 K to 2600 K (or even to the surface temperature of approximately 2800 K) requires relatively little energy (see above and Figure~\ref{fig:energy}), expressing the available $\rm H_2$ energy reservoir in terms of heating from 2000 K is justified. 

The presence of the $\rm H_2$ dissociation thermostat is closely linked to the abundance of molecular hydrogen in stellar atmospheres. As we move to stars with higher effective temperatures, the amount of $\rm H_2$ rapidly decreases, simply because the atmospheric temperatures become too high for significant molecular hydrogen to exist. Consequently, the $\rm H_2$ thermostat becomes progressively weaker and eventually disappears altogether (see Figure~\ref{fig:hotter}). This transition occurs around early M-dwarfs and late K-dwarfs, whose atmospheres are already too warm to sustain sufficient $\rm H_2$ for the thermostat mechanism to operate effectively. The temperatures of white-light flares on these stars are regulated by the hydrogen ionization thermostat and can significantly exceed the stellar effective temperature (see right panel of the cartoon in Figure~\ref{fig:cartoon}).

\vspace{2mm}

\section{Conclusions}
{Our study demonstrates that the comparatively low temperatures of white-light flares on TRAPPIST-1 can be naturally explained by an efficient thermostat mechanism operating in the dense, cool lower atmosphere of an ultracool dwarf. Using chemical-equilibrium calculations and isobaric heat capacities, we show that the dissociation of molecular hydrogen provides a dominant energy sink: as flare heating raises the local temperature a large fraction of the input energy is spent on breaking H$_2$ rather than further increasing the temperature. This H$_2$ dissociation thermostat therefore limits the physical temperature of the photospheric layers that produce the slowly decaying optical/infrared flare component to 3500--4000\,K, in contrast to the 9000\,K temperatures characteristic of solar white-light flares.}

{The efficiency of this thermostat is closely tied to the abundance of molecular hydrogen and thus to stellar effective temperature and atmospheric pressure. As a rule of thumb, the H$_2$ thermostat does not operate for stars hotter than about 4000\,K, where the atmosphere is already too warm for significant H$_2$ to survive and even relatively weak flares are expected to reach higher white-light temperatures.}

{Our treatment is deliberately minimalist and is not intended to replace full radiative-hydrodynamic or magnetohydrodynamic flare simulations. We do not model energy transport from the reconnection site, solve the time-dependent hydrodynamics, or perform non-LTE radiative transfer, and we therefore do not attempt to compute detailed flare spectra. Such RHD and MHD simulations will be essential to obtain physically consistent flare spectral energy distributions, which serve as crucial inputs for studies of atmospheric escape, photochemistry, and climate on exoplanets orbiting active stars \citep{exo1, exo2, exo3, exo4}. The H$_2$ thermostat identified here provides a simple thermodynamic constraint for those future models: regardless of the details of the energy transport and radiative transfer, any successful flare simulation for ultracool dwarfs must be able to reproduce  the low continuum temperatures now being inferred from JWST.}

\begin{acknowledgments}
{We thank Ward Howard and the anonymous referee for their valuable and constructive comments.} The study was supported by the
European Research Council (ERC) under the European Union's
Horizon 2020 research and innovation program (grant No.
101118581---project REVEAL).  VV and NK acknowledge support from the Max Planck Society under the grant ``PLATO Science'' and from the German Aerospace Center under ``PLATO Data Center'' grant   50OO1501. This material is based upon work supported by the National Aeronautics and Space Administration under Agreement No.\ 80NSSC21K0593 for the program ``Alien Earths''. The results reported herein benefited from collaborations and/or information exchange within NASA’s Nexus for Exoplanet System Science (NExSS) research coordination network sponsored by NASA’s Science Mission Directorate.
\end{acknowledgments}

\vspace{5mm}

\appendix
\renewcommand{\thefigure}{\thesection.\arabic{figure}}
\setcounter{figure}{0}
\section{The structure of TRAPPIST-1 photosphere}

\begin{figure}
\centering
 \includegraphics[width=0.5\linewidth]{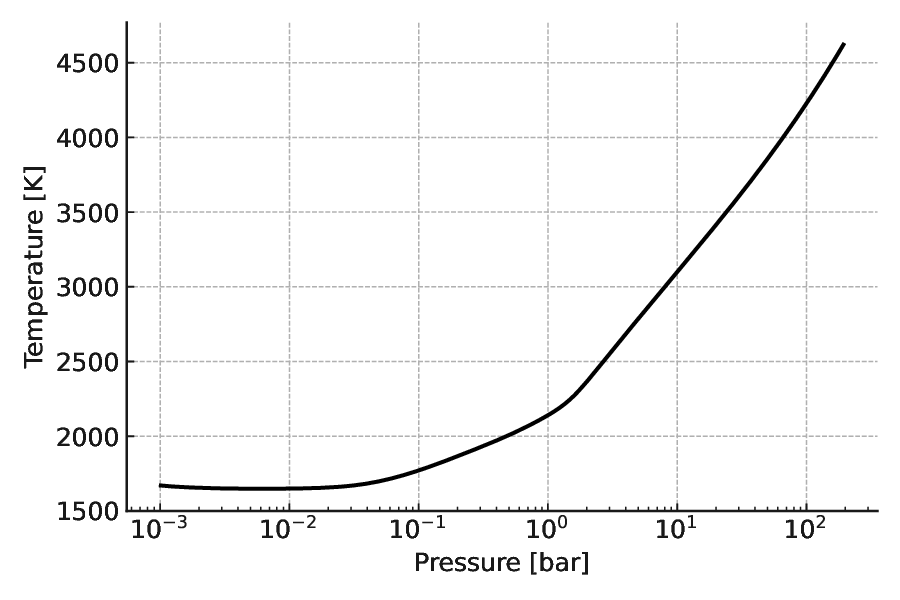}
  \caption{ {\bf The structure of the TRAPPIST-1 photosphere}. Horizontally averaged structure of the 3D photosphere of TRAPPIST-1 computed with the MURaM code. Simulations are performed for an effective temperature $T_{\rm eff}=2580\,\mathrm{K}$, surface gravity $\log g = 5.33$, and solar metallicity. }
   \label{fig:TP_structure}
   \end{figure}

\end{document}